\documentclass[conference]{IEEEtran}
\IEEEoverridecommandlockouts
\usepackage{cite}
\usepackage{amsmath,amssymb,amsfonts}
\usepackage{graphicx}
\usepackage{textcomp}
\usepackage{xcolor}
\usepackage{algpseudocode}
\usepackage{algorithm}
\usepackage{url}
\usepackage{comment}
\usepackage[caption=false, font=footnotesize]{subfig}

\def\BibTeX{{\rm B\kern-.05em{\sc i\kern-.025em b}\kern-.08em
    T\kern-.1667em\lower.7ex\hbox{E}\kern-.125emX}}
\begin{document}

\title{A simulation of urban incidents involving pedestrians and vehicles based on Weighted A*.
}

\author{\IEEEauthorblockN{Edgar Gonzalez Fernandez}
\IEEEauthorblockA{\textit{INFOTEC Centro de Investigación e Innovación en TIC}\\
Aguascalientes, Mexico\\
edgar.gonzalezf@infotec.mx, edgar.gonzalez@ucm.es\\ORCID: 0000-0002-1176-2057\\}
}

\maketitle

\begin{abstract}
This document presents a comprehensive simulation framework designed to model urban incidents involving pedestrians and vehicles. Using a multiagent systems approach, two types of agents—pedestrians and vehicles—are introduced within a 2D grid-based urban environment. The environment encodes streets, sidewalks, buildings, zebra crossings, and obstacles such as potholes and infrastructure elements. Each agent employs a weighted A$^*$ algorithm for pathfinding, allowing for variation in decision-making behavior such as reckless movement or strict rule-following. The model aims to simulate interactions, assess risk of collisions, and evaluate efficiency under varying environmental and behavioral conditions. Experimental results explore how factors like obstacle density, presence of traffic control mechanisms, and behavioral deviations affect safety and travel efficiency.
\end{abstract}

\begin{IEEEkeywords}
Intelligent Agents, Pathfinding, Urban Simulation, Weighted A*
\end{IEEEkeywords}

\section{Introduction}
Urban environments are complex systems where pedestrians and vehicles coexist within constrained and dynamic spaces. Modeling these environments is essential for understanding factors contributing to traffic congestion, pedestrian safety, and infrastructure effectiveness. For instance, official statistics on urban and suburban road crashes in Mexico are compiled by INEGI’s ATUS program~\cite{inegi2023atus} and provide disaggregated series back to the late 1990s, supporting city and state-level risk assessments and trend analysis. At the national level, the 2024 ITF–OECD Road Safety Country Profile~\cite{itf2024mexico} reports that total road deaths decreased by about 6.6\% (2012–2022) despite rapid motorization, while the user-type composition has shifted, underscoring persistent vulnerability for non-motorized users.

In Mexico City, administrative data from the Secretaría de Movilidad (SEMOVI)~\cite{semovi1q2025} highlight the persistent vulnerability of pedestrians and motorcyclists. The most recent quarterly traffic accident report (January–March 2025) indicates that pedestrians consistently account for between 32\% and 42\% of casualties, while the share of motorbike-related casualties has shown a steady increase across the first quarters of 2019 through 2025. This reinforces the need for targeted pedestrian protections and infrastructure improvements.

\begin{figure}[!h]
    \centering
    \subfloat[]{
    \includegraphics[width=.8\linewidth]{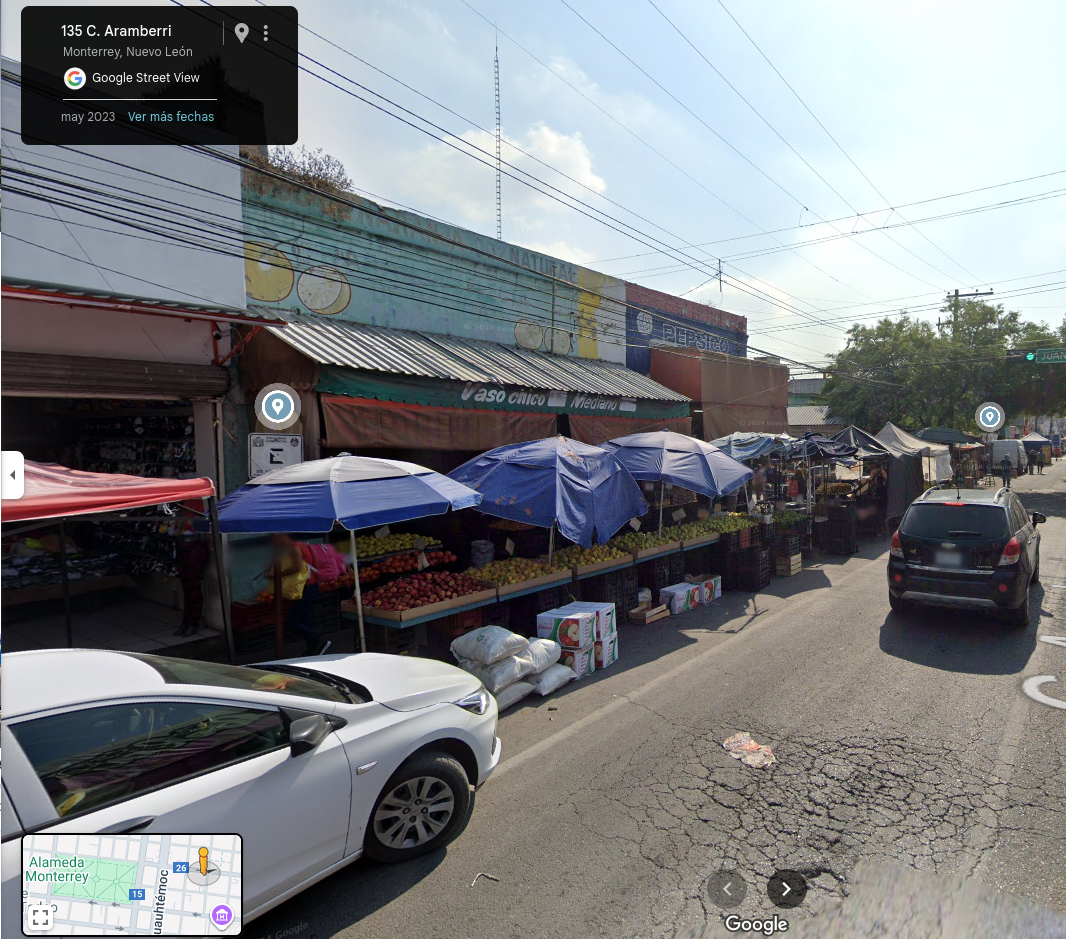}}
    
    \subfloat[]{
        \includegraphics[width=.8\linewidth]{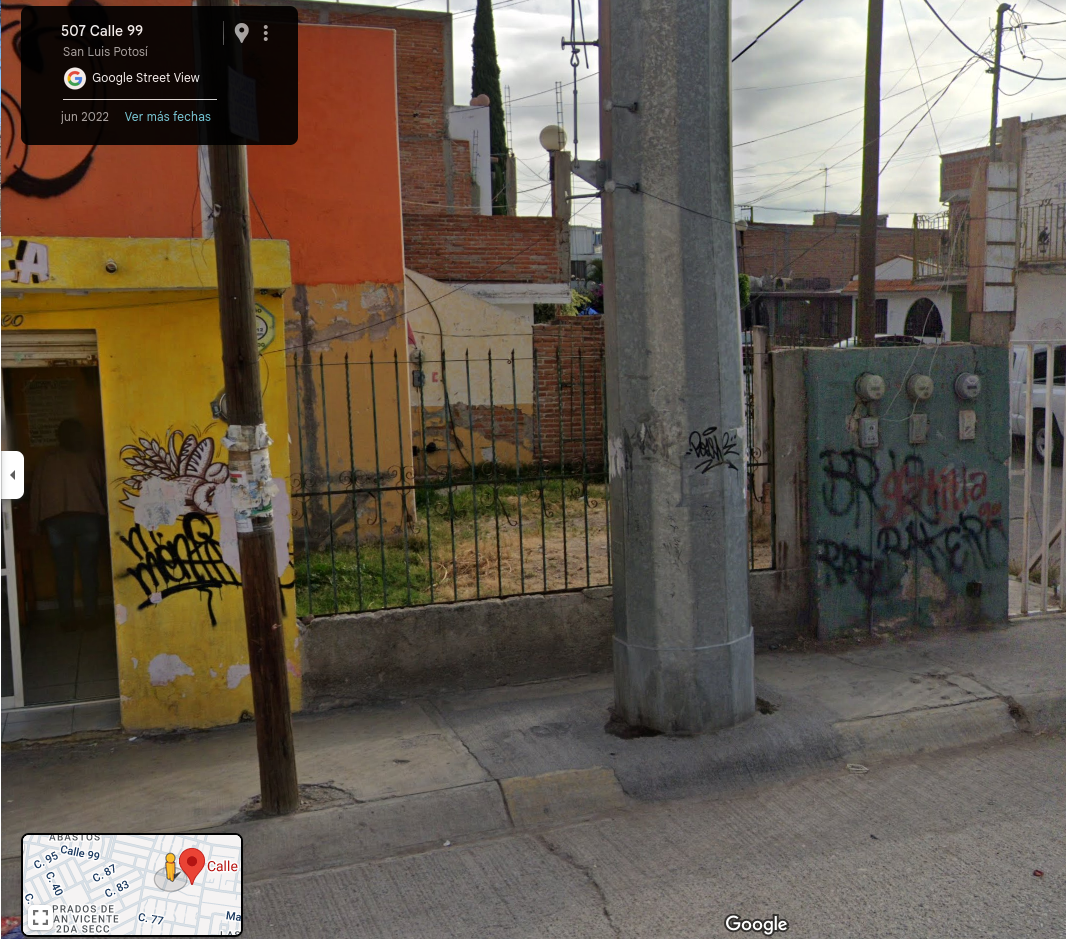}}
        
    \caption{Examples of pedestrian obstacles: (a) a local market, (b) public infrastructure. These elements force pedestrians to walk on the street.\\\scriptsize{Source: Google Maps~(\protect\url{maps.google.com})}.}
    \label{fig:pedestrian_obstacles}
\end{figure}


It is a common problem in Latin American countries to have obstacles difficulting and increasing pedestrian risks (Fig.~\ref{fig:pedestrian_obstacles}). Sidewalks are often narrow, intermittently interrupted by driveways, utility poles, or informal vendors, making them hazardous and inconvenient for pedestrians. In a recent analysis, nearly a third of urban blocks in Mexico were found to lack full sidewalk coverage~\cite{froehlich2020sidewalks}. In Mexico City specifically, a staggering 91.3\% of pedestrian crossings were rated unsafe, disproportionately affecting accessibility and safety for walkers~\cite{paho2021pedestrian}. This inadequate infrastructure forces many pedestrians to walk in the street, significantly increasing their risk of accidents.

The primary goal of the proposed solution and implementation is to develop a precise and adaptable simulation model capable of capturing and analyzing traffic dynamics, with a particular focus on identifying conditions that may lead to pedestrian casualties. By modeling the complex interactions between pedestrians, vehicles, and environmental factors, the system aims to provide reliable insights into risk-prone situations and their underlying causes. Such a model holds significant potential for real-world applicability, as it can be integrated with real data from urban monitoring systems to evaluate current mobility conditions, forecast hazardous scenarios, and assess the effectiveness of proposed interventions. Ultimately, this approach seeks to bridge the gap between theoretical analysis and practical urban planning, offering a decision-support tool that can inform policy-making and infrastructure design to enhance both traffic efficiency and pedestrian safety.

The remainder of this document is structured as follows. Section 2 summarizes related techniques and relevant implementations in the context of urban mobility modeling. Section 3 presents the design of the multiagent system, detailing agent definitions, behaviors, rules, and interaction mechanisms. Section 4 describes the implementation details and the experimental setup used for the simulations. Section 5 reports the results obtained across a wide range of parameter configurations and examines an additional scenario inspired by real-world conditions. Finally, Sections 6 and 7 provide the conclusions of this work and outline directions for future research.

\section{Related Work}

According to their objectives, traffic models are traditionally classified into three main categories~\cite{algherbal2025comparative}:
\begin{itemize}
    \item \textbf{Macroscopic models} analyze traffic as a continuous flow and focus on aggregate measures such as density, flow, and speed. These models are efficient and scalable, making them suitable for citywide or regional planning, though they simplify or ignore individual behavior. These models typically rely on graph-based network representations, hydrodynamic flow equations, and continuum formulations such as the Lighthill–Whitham–Richards (LWR) model~\cite{10.1098/rspa.1955.0089,Richards1956ShockWO} and the Payne–Whitham model~\cite{Payne1971MODELSOF}.
    \item \textbf{Microscopic models} simulate the behavior of individual agents using explicit rules for movement, perception, and interaction. These models capture detailed, safety-critical phenomena such as jaywalking, collisions, pedestrian bottlenecks, and lane-changing behavior, although their higher computational cost limits scalability. Widely used examples include the Intelligent Driver Model (IDM)~\cite{treiber2013traffic} for vehicle dynamics and the Social Force Model~\cite{helbing1995socialforce} for pedestrian movement, both of which allow fine-grained analysis of local interactions and emergent patterns.
    \item \textbf{Mesoscopic models} provide an intermediate level of detail by representing agents as discrete units while simplifying their interactions and movement rules. This balance allows key metrics—such as queuing behavior, travel times, and congestion propagation—to be captured with reasonable realism and computational efficiency. Mesoscopic approaches commonly employ queue-based models, cell-transmission models (CTM)~\cite{alma9976486744202441}, and gas-kinetic traffic formulations~\cite{Hoogendoorn:2001}, making them suitable for evaluating urban policies such as congestion pricing, dynamic tolling, or route management strategies.
\end{itemize}

In multi-agent urban simulations, pathfinding plays a central role in representing how agents move across networks. Classical A* is widely used to compute optimal paths with admissible heuristics such as Euclidean or Manhattan distance~\cite{karur2021pathplanning}. Weighted A* extends this approach by allowing a trade-off between computational efficiency and path optimality~\cite{pohl1970heuristic,likhachev2004ara}. making it more suitable for real-time or large-scale applications. Dynamic environments pose further challenges, requiring algorithms such as D$^*$\cite{stentz1994dstar} and LPA$^*$~\cite{koenig2004lpa} to support replanning when obstacles appear.For multi-agent settings, methods like Conflict-Based Search (CBS)~\cite{sharon2015conflict}, Optimal Reciprocal Collision Avoidance (ORCA)~\cite{berg2008orca}, and reinforcement learning\cite{martinez2011rlpedestrian}have been introduced to coordinate multiple agents and avoid conflicts.

Beyond pathfinding, behavioral models add realism to how agents interact locally. Pedestrian movement is represented as a combination of attractive and repulsive forces by SFM, capturing collision avoidance and crowd effects. For vehicles, IDM~\cite{treiber2013traffic} has become a standard for modeling acceleration, braking, and car-following behavior. By combining global planning methods such as A$^*$ with behavioral models like SFM and IDM, simulations can reproduce both large-scale mobility patterns and fine-grained, agent-level interactions, which is particularly important for mixed pedestrian–vehicle environments.

In this work, we introduce Weighted A$^*$ into multi-agent urban simulations not only as a computational tool but also as a behavioral modeling mechanism. By adjusting weights, agents can exhibit a range of strategies: from lawful and safe movements to opportunistic or reckless actions such as jaywalking or illegal turns. This dual role of Weighted A* balances efficiency with behavioral realism, positioning it as a promising method for investigating how infrastructure design and heterogeneous decision-making interact to shape urban mobility. Our aim is to demonstrate its potential to generate realistic, data-driven insights that bridge computational efficiency with social complexity.



\section{Multiagent system design}
\label{MASDesign}

This section presents the design and implementation of the proposed multi-agent simulation. The description is organized around three key elements: the environment model, agent definitions, and interaction mechanisms. For this task, we have considered pedestrians and vehicles as agents interacting in the city grid, each provided with different behavior, sensing capabilities, and decision-making strategies. Their interactions include collision avoidance, right-of-way adherence, and cooperative maneuvers. The grid-based encoding contains structured information interpretable by agents at each cell, enabling algorithms like Weighted A* to operate appropriately. Together, these components establish a coherent framework for simulating and analyzing complex mobility patterns within urban environments.

\subsection{Environment}
The environment is designed to replicate a city grid composed of roads, sidewalks, pedestrian crossings, and obstacles. It is represented as a two-dimensional grid in which each cell encodes ground type and permitted traffic directions using a three-character string. The first character specifies the cell type: road (r), sidewalk (s), building (b), parking (p), or zebra crossing (z). The second and third characters denote the allowed traffic flow directions for vehicles: North (N), South (S), East (E), and West (W). This encoding makes it possible to define both circulation rules and turning options at intersections for basic but common driveway directions.

\begin{figure}[h!]
    \centering
        \includegraphics[width=.9\linewidth]{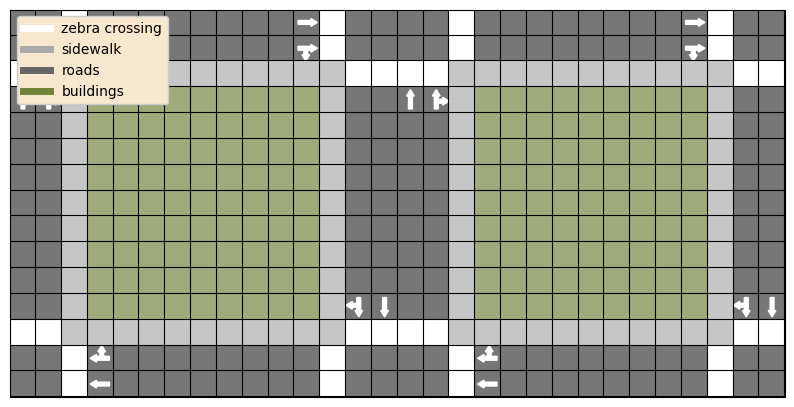}
    \caption{Image shows codification of a basic environment. Some arrows have been added to observe directions and cells that enforce turns for driver agents.}
    \label{fig:design_environment}
\end{figure}

A visualization of the encoding for a basic urban environment consisting of two blocks and 4 lanes (2 for each direction) is shown in Fig.~\ref{fig:design_environment}. White arrows highlight the allowed directions for cells in the central square, which are designated as preferred positions for vehicles when executing turns. Temporary elements such as obstacles, potholes, or damaged infrastructure can be incorporated into the environment by listing their coordinates, enabling the simulation to capture dynamic and realistic urban conditions.

\subsection{Agents}
Agents are the active entities within the simulation, representing both pedestrians and vehicles, each with defined goals, sensing capabilities, and behavioral strategies. Their design enables the modeling of heterogeneous decision-making processes and agent interactions, providing valuable insights into how individual actions collectively shape urban mobility patterns.

The agentic model aims to reproduce realistic mobility dynamics through decision-making processes driven by pathfinding and cost optimization. Each autonomous agent (walker or driver) plans its trajectory using a Weighted A* algorithm, which minimizes a combined cost–risk function balancing travel efficiency and behavioral realism. 

The cost function $g(v)$ of a cell $v$ defines the inherent risk of each type of cell in the grid, while the risk function $r(a, v)$ introduces behavioral variability, defined for each pair of action-position, assigning different values to each possible action according to the current agent's position. This will be particularly useful to enforce driver behavior and adherence to traffic rules.

Formally, the function to be applied for the modified weighted A$^*$ search is as follows: 

\begin{equation}
    f(n) = g(v) + w h(v) + \alpha r(a, v).
    \label{eq:weighted_astar}
\end{equation}
where $g(v)$ is the cost of visiting cell $v$, $h(v)$ the selected heuristic (Manhattan distance) from cell $v$ to the goal, $r(a, v)$ the risk of performing action $a$ in cell $v$, and $w, \alpha$ weights to model agent's individual behavior. The explicit and detailed algorithm can be consulted in Appendix A.

\subsubsection{Walker agents (pedestrians)}

 Walker agents simulate pedestrians traveling from one building to another. Their main characteristics are as follows:
    \begin{itemize}
        \item They plan routes originating from building cells and navigate primarily via sidewalks and zebra crossings, although road cells may be used when necessary or preferred.
        \item Pathfinding is performed using a weighted A$^*$ algorithm, where higher weights prioritize more direct paths, but at the cost of favoring unsafe actions, such as jaywalking.
        \item Upon detecting an approaching vehicle, walkers may either stop or reroute, depending on whether the vehicle is active or inactive (e.g. parked or damaged).
        \item Walkers reaching their goals are removed from the environment.
    \end{itemize}
\begin{table}[!h]
  \begin{tabular}{|l|c|l|c|}
     \hline
    \textbf{Cell type}  &\textbf{Symb.} &\textbf{Description}    &\textbf{Cost}\\
     \hline
     \scriptsize
     Sidewalk   &s      &Designated pedestrian area &1\\ 
     Zebra crossing   &z      &Authorized pedestrian area &1\\ 
     Road   &r      &Vehicle lanes &5\\ 
     Intersections (turn cells)   &t,l      &Vehicle turning area &10\\
     Building/Obstacles   &b, o      &Impassable element &$\infty$\\
     \hline
  \end{tabular}
  \caption{Cost function for walker agents.}
  \label{tab:walker_costs}
\end{table}

For these agents, a cost function reflects pedestrian accessibility and safety priorities. Movement through sidewalks and zebra crossings incurs a minimal travel cost, as these are safe and designated pedestrian zones. Roads are assigned a higher cost, discouraging walking on vehicle lanes except in constrained conditions. Intersections, where vehicles may turn, are considered to represent elevated risk. Obstacles and buildings are assigned infinite (or very large) costs, preventing traversal. For pedestrians, the risk function remains zero in all cases, as their behavior is assumed lawful and risk-averse. Table~\ref{tab:walker_costs} summarizes example values for pedestrian costs used in simulations. Higher costs are associated with riskier areas.

\subsubsection{Driver agents (vehicles)}

Driver agents are enforced to travel on road cells, and plan a route preferring a path using the direction encoded by the cell. A higher weight can result in a path that travels against the designated road direction. Briefly, drivers stick to the following rules: 
    \begin{itemize}
        \item They start a journey from road ends or buildings, and travel to either exit the map, reach a building entry, or park in allowed spots.
        \item They must follow traffic rules, including directional movement, stop at lights, and yielding. Some turns may be decided to arrive to parking places.
        \item Drivers reaching their goals are removed; those reaching designated parking places remain there until reactivated (e.g., when a new goal is assigned).
    \end{itemize}
\begin{table}[h!]
  \begin{tabular}{|l|c|l|c|}
    \hline
    \textbf{Cell type}  &\textbf{Symb.} &\textbf{Description}    &\textbf{Cost}\\
    \hline
    Road   &r      &Vehicle lane &1\\ 
    Zebra crossing   &z      &Pedestrian area &1\\ 
    Parking space   &p      &Parking area &5\\
    Pothole   &h      &Vehicle lanes &5\\
    Sidewalk/Building/Obstacles   &s, b, o      &Impassable element &$\infty$\\
    \hline
  \end{tabular}
  \caption{Cost function for driver agents.}
  \label{tab:driver_costs}
\end{table}
The cost function prioritizes efficient roadway traversal while penalizing non-ideal choices. Table~\ref{tab:driver_costs} summarizes example values for the cost function according to cell type (roads, zebra-crosses, and exclusive pedestrian cells), giving preference to exclusive vehicle ground and minimizing unnecessary turns or visiting damaged infrastructure. Zebra crossings and parking spaces have been assigned a bigger value to minimize crossing paths where pedestrian or vehicle interaction is higher.
\begin{table}[h!]
  \begin{tabular}{|l|l|c|}
    \hline
    \textbf{Action} &\textbf{Description}    &\textbf{Risk}\\
    \hline
    Forward   &Movement on road direction &0\\ 
    Right turn   &Legal right turn &1\\ 
    Left turn   &Legal left turn      &2\\
    Lane change   &Horizontal lane change      &3\\
    Invalid turn   &Illegal turn& 5\\
    Backward/wrong-direction   &Irresponsible or illegal movement &20\\
    \hline
  \end{tabular}
  \caption{Risk function for driver agents.}
  \label{tab:driver_risks}
\end{table}

The risk function for drivers introduces behavioral heterogeneity to simulate varying degrees of lawfulness and recklessness. Safe maneuvers, such as forward movement, have no penalty. However, more hazardous or even illegal actions, such as left or right turns in non-valid cells, or backward and wrong-direction driving, add higher penalties. Tables~\ref{tab:driver_costs} and~\ref{tab:driver_risks} exemplify appropriate values to discourage reckless driving.

\begin{figure}[h!]
    \centering
    \subfloat[]{
        \includegraphics[width=0.4\textwidth]{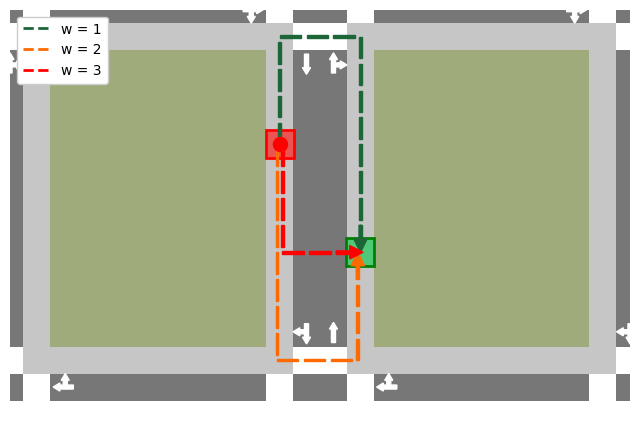}}
    
    \subfloat[]{
        \includegraphics[width=0.4\textwidth]{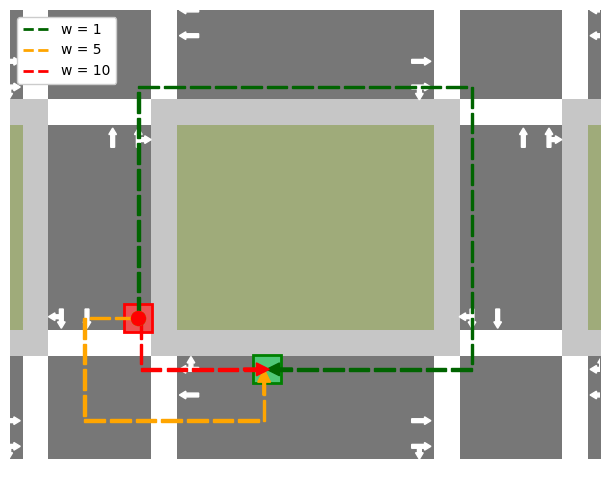}
    }
    \caption{Routes obtained using Weighted A$^*$ with different weights for identical start and goal positions. Start and goal cells are highlighted in red and green, respectively. (a) Pedestrian routes showing instances of jaywalking for $w = 5$. (b) Driver routes illustrating reckless behavior, including opposite-direction travel, for $w = 10$.}
    \label{fig:astar_agents}
\end{figure}

Both types of agents will interact in the environment according to their own rationale. Fig.~\ref{fig:astar_agents} illustrates the expected behavior of agents with different weights, showing how these parameters influence the decision of agents regarding the best route.

\subsection{Reasoning and Interactions}
The simulation operates in discrete time steps, where each agent continuously perceives, reacts to, and acts within a dynamic environment. Upon initialization, a new agent plans a route using the Weighted A$^*$ algorithm, parameterized according to its assigned behavioral weights as defined in Eq.~\ref{eq:weighted_astar}. The heuristic $h(v)$ is computed as the Manhattan distance between the current cell 
$v$ and the goal, combined with the risk penalty $r(a, v)$ proportional to the potential danger associated with traversing that cell under a certain action. This allows agents to optimize their movement based on both efficiency and safety considerations.

\begin{figure}[h!]
    \centering
    \includegraphics[width=.9\linewidth]{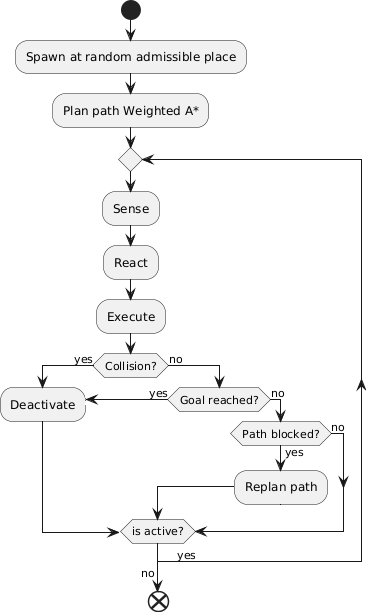}
    \caption{General flowchart of agent implementation. Agents are deactivated immediately upon reaching their goal, or with a delay in the event of a collision. For vehicles, parking results in an inactive state at the same position until a new goal is assigned.}
    \label{fig:agent_diagram}
\end{figure}

At every simulation step, agents execute a structured routine composed of four phases: sensing, reacting, acting, and iterating, as shown in~\ref{fig:agent_diagram}.
\begin{enumerate}
    \item \textbf{Sensing}. Each agent perceives its surroundings within a fixed local neighborhood, set to an area that encompasses several future steps along its intended route. This local sensing mechanism allows agents to anticipate potential collisions or the presence of dynamic obstacles, such as other agents or temporary objects. Collision detection relies on the Euclidean distance between entities: two vehicles are considered in conflict if their distance is less than one cell unit, whereas pedestrians, represented with a smaller effective radius of 0.1 units, can occupy the same cell concurrently to simulate realistic crowd density. Vehicle agents additionally align their position to predefined lane centers to ensure proper road alignment and lane discipline. 
    \item \textbf{Reacting}. Collision detection ensures avoidance protocols, where agents delay or reroute when necessary:
    \begin{itemize}
        \item Pedestrians stop immediately if a potential collision is detected. However, on a zebra crossing, they assume right-of-way and proceed without delay, unless an obstacle is in their way.
        \item Drivers monitor their distance to other agents, decelerating when necessary and accelerating up to their maximum speed otherwise. At zebra crossings, they yield to nearby pedestrians, including those on sidewalks. However, on driveways, they maintain speed unless a pedestrian steps into their path. A sudden pedestrian invasion may lead to a collision.
        \item Both agent types may re-plan their route if they encounter unforeseen obstacles, such as recently damaged agents, that were not detected during the initial path planning.
    \end{itemize}
    \item \textbf{Acting}. Following the sensing and reaction phases, agents update their intended action based on the adjusted plan and execute it from their current cell. This may involve advancing to the next position, stopping, or modifying orientation depending on the contextual conditions and newly computed trajectory.
    \item \textbf{Iterating}. A global scheduler manages all updates to ensure consistent simulation timing and agent coordination. It handles the removal of damaged or deactivated agents, reactivation of parked vehicles when assigned new destinations, and instantiation of new agents according to predefined arrival rates or environmental conditions. This centralized control maintains synchronicity among all entities and supports dynamic environmental changes throughout the simulation.
\end{enumerate}

\begin{figure}[!h]
    \centering
    \subfloat[]{\includegraphics[width=.47\linewidth]{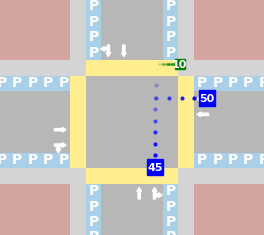}}\hfill
    \subfloat[]{\includegraphics[width=.47\linewidth]{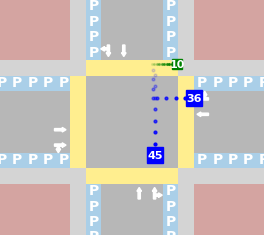}}
    
    \subfloat[]{\includegraphics[width=.47\linewidth]{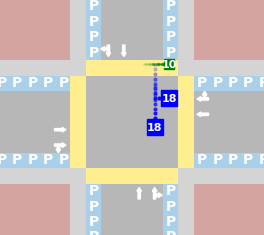}}\hfill
    \subfloat[]{\includegraphics[width=.47\linewidth]{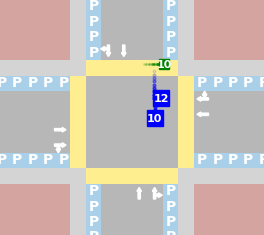}}
    
    \caption{Sequence of frames showing the reactions of different agents. The green pedestrian walks through a zebra crossing and continues along the path without modifying its actions, while the blue vehicles react by reducing their speed (white number) to avoid a collision or a potential runover. Small dots indicate the next position of the agent according to current attributes (speed and direction).}
    \label{fig:agent_reaction}
\end{figure}

Figure~\ref{fig:agent_reaction} shows how some of these rules are observed on a simulation sequence. Runovers or collisions may happen in scenarios where little time is left for an effective reaction, for example, when pedestrians step down to the road when a driver is near.

\subsection{Implemenation details}
The simulation was implemented in Python using the AgentPy framework, which provides a flexible structure for defining agents, environments, and interaction mechanisms in multi-agent modeling. The implementation focuses on creating a configurable and extensible system capable of representing diverse urban mobility conditions. Several behavioral parameters were designed to allow controlled experimentation and to reproduce realistic city dynamics.

The simulation enables customization of the number of agents present in the environment, including both pedestrians and vehicles. This parameter allows users to test scenarios ranging from light traffic to highly congested street conditions. AgentPy’s population tools were used to initialize these agents and manage their lifecycle throughout the simulation.

To reproduce realistic variations in urban settings, spawning functions were implemented to introduce agents according to different behavioral or structural assumptions. For example, Poisson-distributed spawning was used to model spontaneous arrivals of pedestrians and vehicles, while controlled spawning densities were applied to simulate street markets or temporarily crowded zones.

Agent behavior can be diversified by individually assigning a variable maximum speed and risk factor to both pedestrians and drivers. These values could be set as constants or sampled from probabilistic distributions to reflect cautious drivers, elderly pedestrians, or individuals with mobility impairments. Incorporating such heterogeneity allows the model to capture more realistic and heterogeneous mobility patterns.

Finally, the environment incorporates randomly positioned or predefined obstacles representing elements such as potholes, construction barriers, or sidewalk obstructions. These obstacles influence routing decisions and alter agent trajectories, enabling the study of how infrastructure limitations affect safety and mobility outcomes within the simulated environment.


\section{Experimental configuration}

The experimental setup is designed to evaluate how variations in the urban environment influence agent behavior and overall mobility performance. We focus on pedestrian safety and traffic efficiency, with a particular interest in the effects of sidewalk obstacles. Pedestrians and vehicles are represented as autonomous agents with decision-making strategies and interaction rules as previously defined. For controlled observation, we vary the number and distribution of sidewalk obstacles, which act as constraints that may force pedestrians to deviate from safe walking areas, increasing potential risk. In parallel, vehicle average speed is monitored to capture how pedestrian detours and congestion points impact overall traffic flow.

Table \ref{tab:exp_parameters} summarizes the experimental conditions used to evaluate the proposed multiagent simulation framework. The environment consists of a 5×5 grid of urban blocks, each measuring 15×15 cells. Streets surround the perimeter of each block, sidewalks border each street, and the interior 13×13 area represents a building or non-traversable space. This configuration provides a structured yet flexible setting to systematically observe the effects of density, congestion, and obstruction on agent behavior.

To explore a range of realistic urban mobility conditions, three key parameters were varied across multiple executions:

\begin{itemize}
    \item Pedestrian population, ranging from 0 to 200 in increments of 25.
    \item Driver population, ranging from 0 to 100 in increments of 20.
    \item Sidewalk obstacles, randomly placed to block pedestrian flow, at 0\%, 5\%, and 10\% of total sidewalk cells.
\end{itemize}

Simulations conducted without drivers make it possible to isolate the influence of obstacles on pedestrian behavior, highlighting how varying risk weights contribute to the emergence of jaywalking. Conversely, simulations without pedestrians focus exclusively on vehicle dynamics, allowing the analysis of traffic flow patterns driven solely by driver interactions.
\begin{table}[h!]
    \centering
    \begin{tabular}{|l|c|} 
        \hline \textbf{Parameter} & \textbf{Values} \\ 
        \hline \hline Simulation length & 1000 steps \\
        \hline Number of pedestrians & 0, 25, 50, ..., 200 \\
        \hline Number of drivers & 20, 40, 60, 80, 100 \\
        \hline Sidewalk obstruction levels & 0\%, 5\%, 10\% \\
        \hline Urban layout & 5×5 blocks, each 15×15 cells \\
        \hline Building footprint & 13×13 central area per block \\
        \hline Agent replenishment & Yes (population maintained) \\
        \hline Metrics collected & \begin{tabular}{@{}l@{}} Avg. driver speed, \\ Jaywalking count, \\ Runover incidents \end{tabular} \\
        \hline Spatial analysis tools & Heatmaps of conflict zones \\
        \hline \end{tabular}
    \caption{Summary of experimental parameters used in the simulation.} \label{tab:exp_parameters}
\end{table}

These experiments were run for 1000 simulation steps, during which new agents were introduced dynamically into the environment whenever existing agents reached their goals or were removed. This approach maintains approximately constant population levels, ensuring a stable flow of interactions throughout the simulation horizon.

The following performance indicators were collected for analysis:

\begin{itemize}
    \item Average driver speed, representing vehicular mobility efficiency. 
    \item Number of jaywalking events, counted whenever pedestrians step onto road cells.
    \item Number of collisions between vehicles (driver agents).
    \item Number of runover incidents, defined as any collision between a pedestrian and a vehicle.
\end{itemize}

Collisions are determined based on the physical size assigned to each agent type. Vehicles are modeled as occupying 80\% of a cell, while pedestrians occupy 10\%. Thus, vehicles have an effective size of $0.8\times 0.8$ units, and pedestrians $
0.1\times 0.1$ units. A collision is detected whenever the Euclidean distance between two agents becomes smaller than the sum of their effective radii. Up to now, pedestrian collisions have not been considered.

Additionally, spatial heatmaps were generated to identify conflict-prone areas, indicating where jaywalking, slowdowns, or vehicle–pedestrian interactions tend to cluster within the city grid. These visualizations enable a qualitative examination of how infrastructure design, pedestrian obstruction, and traffic density contribute to emerging urban mobility risks.

\section{Results and analysis}
This section presents the outcomes of the simulation experiments and analyzes how pedestrian and vehicle agents behave under varying traffic conditions, environmental configurations, and behavioral weight settings. By examining a range of performance indicators related to travel efficiency and risky events, we assess the impact of infrastructure quality, sidewalk obstruction, and agent decision-making on urban mobility. The experiments also highlight the emergent dynamics of the multiagent system, providing insight into critical zones and behaviors that arise from interactions between walkers, drivers, and the environment. 
\subsection{Pedestrian safety indicators}
\begin{figure*}[!h]
    \centering
    \subfloat[$w=1$]   {\includegraphics[height=4.5cm]{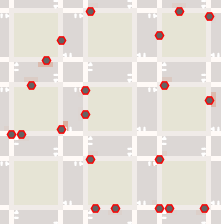}} \hfill
    \subfloat[$w=3$]   {\includegraphics[height=4.5cm]{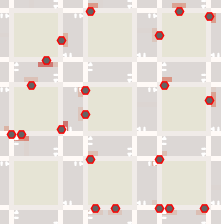}}   \hfill
    \subfloat[$w=5$]   {\includegraphics[height=4.5cm]{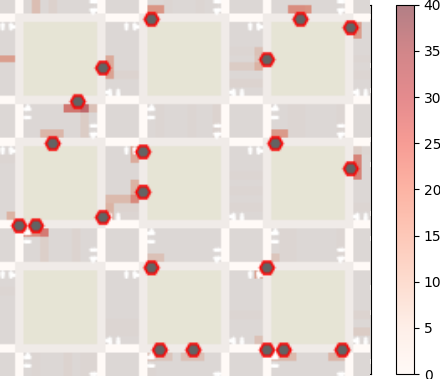}}
    
    \subfloat[]   {\includegraphics[height=5cm]{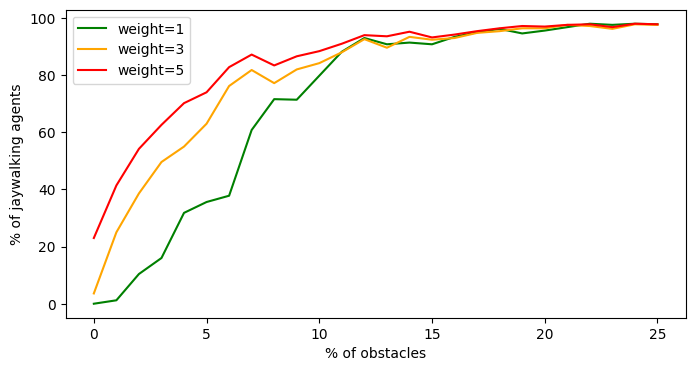}}
    \caption{Heatmaps of city jaywalking with 5\% of sidewalk cells occupied. (a), (b), and (c) illustrate the areas traversed by pedestrian agents with weights 1, 3, and 5, respectively. In (d), the graph shows how an increasing number of obstacles compels pedestrians to enter risky areas more frequently, even when exhibiting risk-averse behavior.}
    \label{fig:heatmaps_jaywalking}
\end{figure*}

    

The first set of experiments focuses on analyzing pedestrian safety and behavior modeling and its interaction with the surrounding environment. Several simulation runs were conducted using different percentages of sidewalk obstacles, randomly distributed across the grid, and varying weight parameters representing the agents’ tendency toward risky or conservative behavior.
Figure~\ref{fig:heatmaps_jaywalking} presents selected results from these experiments. Subfigures (a), (b), and (c) present heatmaps of pedestrian trajectories for weights 1, 3, and 5, respectively. Obstacles are shown as red and grey hexagons, while red shading intensity indicates the frequency with which pedestrians step onto road cells, corresponding to jaywalking events. As the weight increases, pedestrians display markedly riskier behavior, deviating from designated pedestrian facilities such as sidewalks and zebra crossings with greater frequency.

Subfigure (d) provides a quantitative assessment of these trends by reporting the proportion of agents occupying road cells under varying environmental conditions. The percentage of obstructed sidewalk cells is varied from 0\% to 25\%. The results show a sharp increase in jaywalking even under low obstruction levels (approximately 5\%), demonstrating the high sensitivity of pedestrian movement to minor degradations in sidewalk infrastructure and underscoring the importance of maintaining unobstructed pedestrian pathways to ensure safe mobility.

\begin{figure}
\centering    
   \subfloat[]{
        \includegraphics[width=\linewidth]{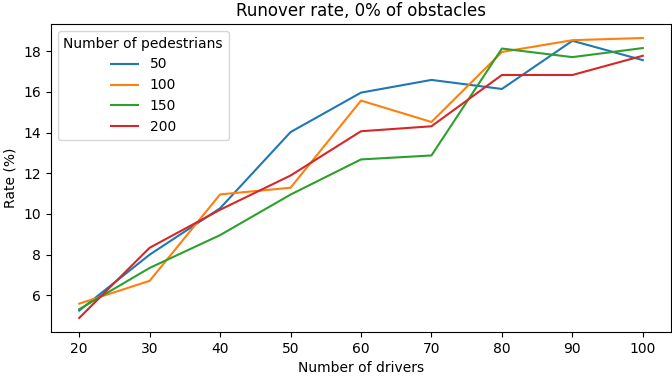}}
   
   \subfloat[]{
        \includegraphics[width=\linewidth]{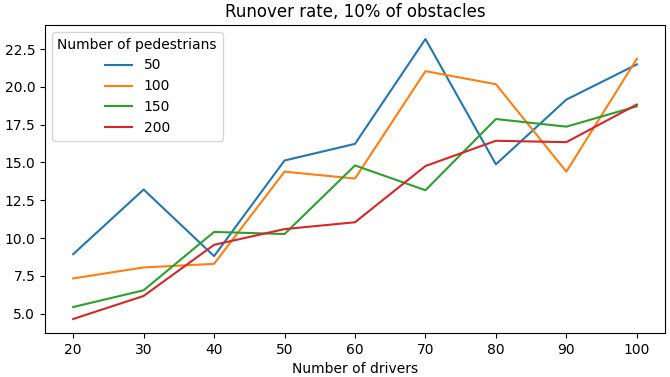}}

    \caption{Graphs showing the trend of pedestrian collision incidents with growing amount of vehicles. In (a) results show incidents considering no obstacles. In (b) we observe the results considering 10\% of sidewalk cells occupied.}
    \label{fig:graph_runover}
\end{figure}

Furthermore, an analysis of runover incidents across different parameter configurations (Fig.~\ref{fig:graph_runover}) reveals two key patterns:
\begin{itemize}
    \item Pedestrian risk increases substantially as the number of vehicles grows, reflecting the higher probability of conflict under denser traffic conditions.
    \item Surprisingly, the number of sidewalk obstacles does not significantly influence pedestrian–vehicle collision rates. This effect may change, however, if vehicle agents are allowed to adopt more aggressive or risk-prone behaviors, suggesting an avenue for deeper exploration in future simulations.
\end{itemize}   

\subsection{Vehicle mobility}
\begin{figure}[!h]
    \centering
    \subfloat[]{
        \includegraphics[width=.75\linewidth]{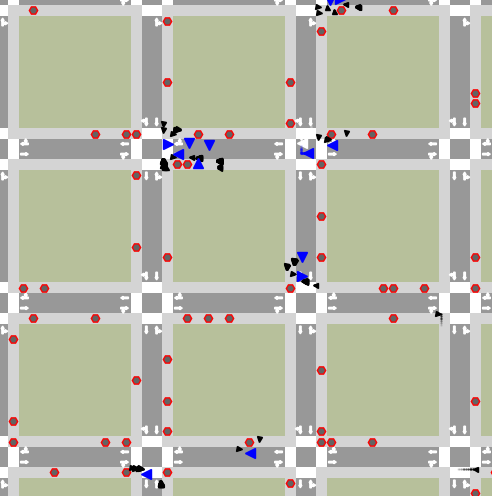}}
    
    \subfloat[]{
        \includegraphics[width=\linewidth]{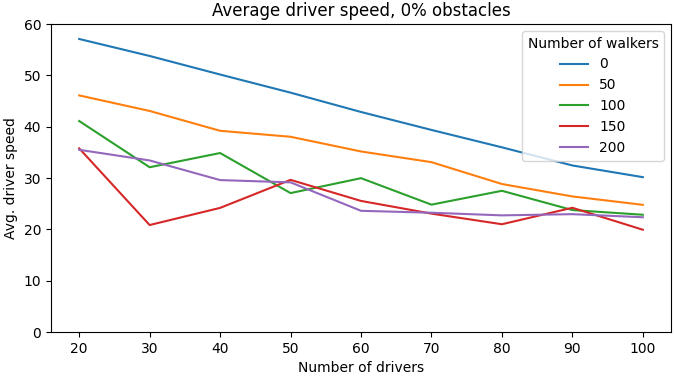}}
    
    \subfloat[]{
        \includegraphics[width=\linewidth]{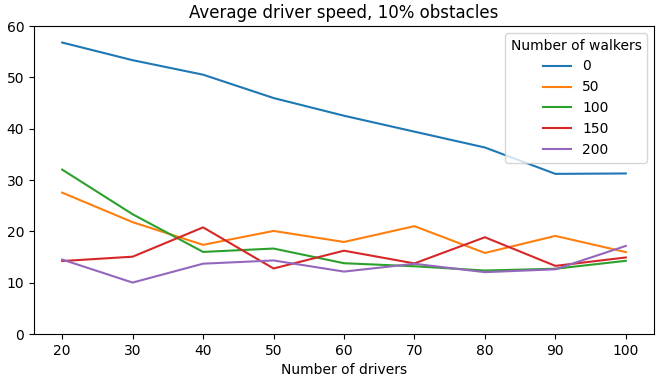}}

   \caption{Effects of agent interactions and obstacle density on traffic flow. (a) Visual depiction of traffic congestion resulting from pedestrians stepping into the roadway due to sidewalk obstacles.
    (b) and (c) Quantitative metrics illustrating how varying numbers of pedestrians and obstacles influence average vehicle speed.}
   \label{fig:graphs_speed}
\end{figure}

This subsection evaluates quantitative mobility indicators for drivers, specifically average vehicle speed and vehicle collision rate, under varying simulation conditions. The analysis examines how mobility performance changes as the number of agents increases and as obstacle density becomes higher. Fig.~\ref{fig:graphs_speed} presents the metrics obtained across multiple simulation runs.

With respect to average vehicle speed, the results indicate a clear dependency on traffic density. As the number of vehicles increases, traffic speed declines rapidly, even in scenarios where no pedestrians are present. Pedestrian density influences vehicle speed primarily in environments with relatively few vehicles; under these conditions, pedestrian movements introduce additional delays and force vehicles to adjust their trajectories. However, once the number of circulating vehicles reaches a sufficiently number (at a maximum of 100 in our tests) the presence of pedestrians no longer produces a substantial effect on the overall average speed. Furthermore, increased obstacle density along sidewalks tends to divert pedestrians into the roadway, which in turn produces moderate reductions in vehicle speed, though this effect is secondary compared to vehicular density itself. In this sense, some simulations have shown heavy traffic after bottlenecks appeared in conflictive areas, as seen in Fig.~\ref{fig:graphs_speed}(a).
\begin{figure}[!h]
\centering    
    \subfloat[]    {    \includegraphics[width=\linewidth]{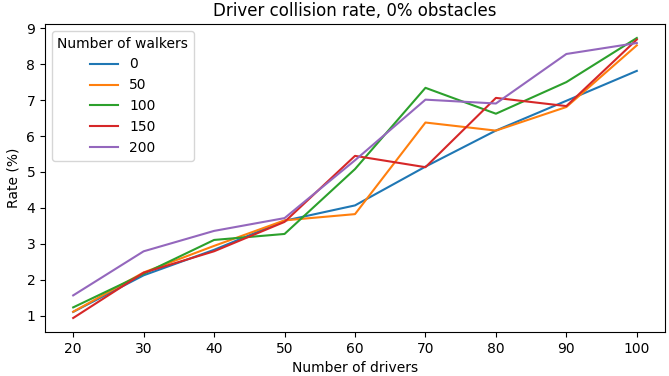}}

    \subfloat[]    {    \includegraphics[width=\linewidth]{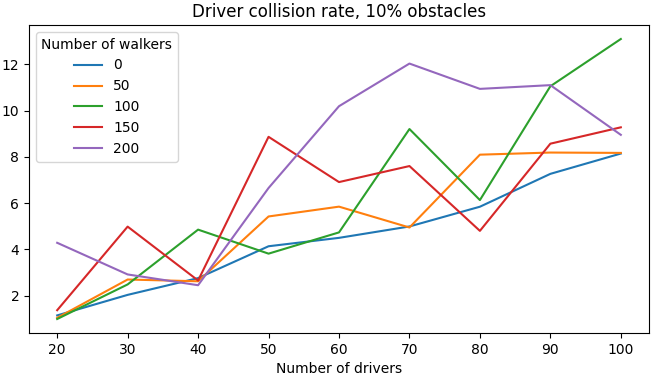}}
    \caption{Graphs showing the influence of vehicular density on collision rates.}
    \label{fig:graphs_collisions}
\end{figure}




The analysis of collision rates reveals similar trends regarding the dominant influence of vehicle density. An increase in the number of vehicles leads to a rapid escalation in collision frequency, even in scenarios that do not include pedestrian agents. Pedestrian density, by contrast, does not produce a significant effect on the collision rate, suggesting that conflicts among vehicles remain the primary source of unsafe interactions within the environment. Obstacle density along pedestrian pathways exerts only a mild influence on collision-related indicators; although obstacles divert pedestrians into traffic lanes, the overall trend in collision behavior is overwhelmingly governed by the number of vehicles present in the simulation rather than by pedestrian or obstacle interactions.

\subsection{Spatial analysis in obstructed urban scenarios}
This subsection analyzes the spatial distribution of mobility conflicts using heatmap visualizations generated from simulations of an urban environment characterized by obstructed sidewalks and high pedestrian density. The experimental scenario represents several city blocks heavily crowded with obstacles, such as those found during a street market event, restricting pedestrian movement and forcing complex interactions with vehicular traffic. This configuration enables observation of the emergent dynamics between drivers and pedestrians under realistic conditions of limited accessibility and environmental clutter.

To introduce behavioral variability, each pedestrian is assigned a randomly sampled risk weight ranging from 1 to 3, while drivers receive risk weights between 1 and 5. These weights influence agents’ willingness to take risky or conservative actions, thereby shaping the resulting spatial patterns of collisions, speed reduction, and local congestion. Table~\ref{tab:usecase_params} summarizes the main elements for the use case simulation.

\begin{table}[h!]
\centering
\begin{tabular}{|l|c|c|}
\hline
\textbf{Parameter} &\textbf{Pedestrians} &\textbf{Drivers}\\
\hline
Agent Risk Weights & Random in [1, 3]& Random in [1, 5] \\
\hline
Number of Agents & 100& 50 \\
\hline
Maximum Agent Speeds & 60 units& 10 units \\
\hline
Population Dynamics & \multicolumn{2}{c|}{Repopulate for a constant population} \\
\hline
Environment Configuration & \multicolumn{2}{c|}{Two street blocks partially obstructed}\\
\hline
\end{tabular}
\caption{Description of parameters used for a simulation in an urban obstructed scenario.}
\label{tab:usecase_params}
\end{table}
Collision frequency reveals concentrated conflict zones near narrow passageways and intersections where pedestrians are forced into the roadway due to sidewalk obstructions. Speed heatmaps show pronounced slowdowns along segments with high pedestrian spillover or frequent interactions between agents of high risk weight. Finally, accumulation heatmaps highlight areas where agents tend to cluster, often forming bottlenecks around obstacles or within constrained sidewalk segments.

By comparing these spatial indicators across multiple runs, the analysis uncovers how both environmental constraints and heterogeneous agent risk profiles contribute to the formation of recurrent hotspots. This spatial perspective complements the quantitative results presented in earlier sections, offering deeper insight into how crowded infrastructure and behavioral variability jointly affect safety, efficiency, and urban mobility performance.
\begin{figure}[!h]
\centering    
    \subfloat[Count of drivers] {
        \includegraphics[width=.48\linewidth]{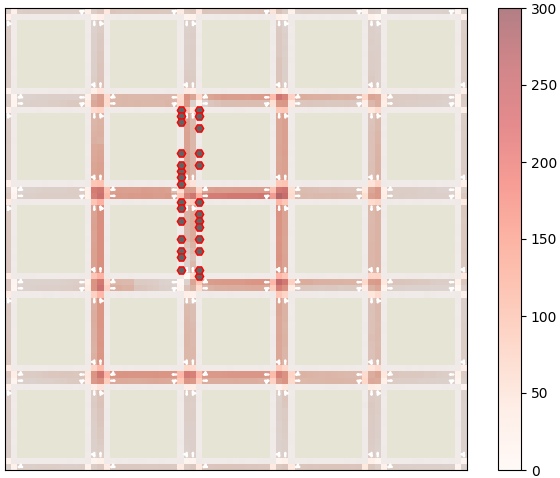}}\hfill
    \subfloat[Average speed of drivers] {
        \includegraphics[width=.48\linewidth]{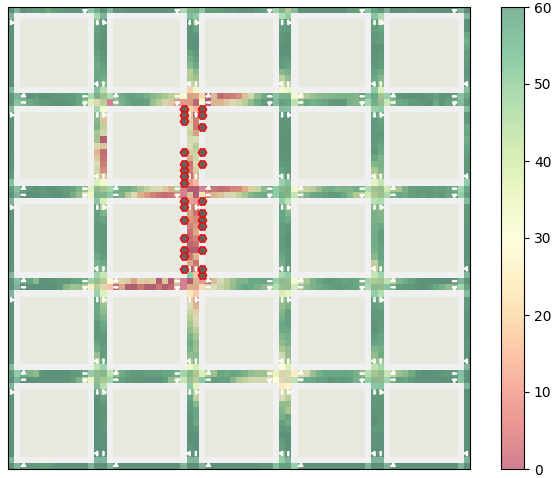}}
        
    \subfloat[Count of pedestrians] {
        \includegraphics[width=.48\linewidth]{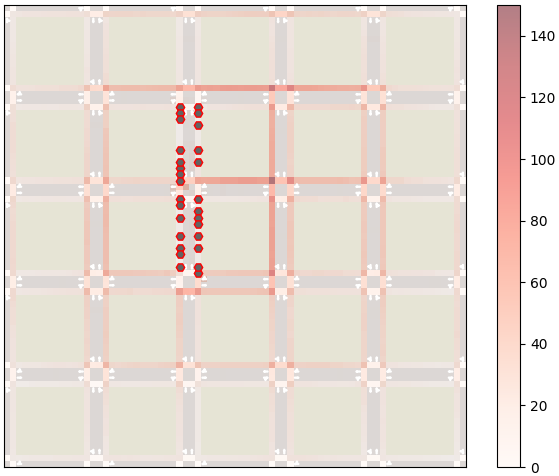}}\hfill
    \subfloat[Count of jaywalking pedestrians] {
        \includegraphics[width=.48\linewidth]{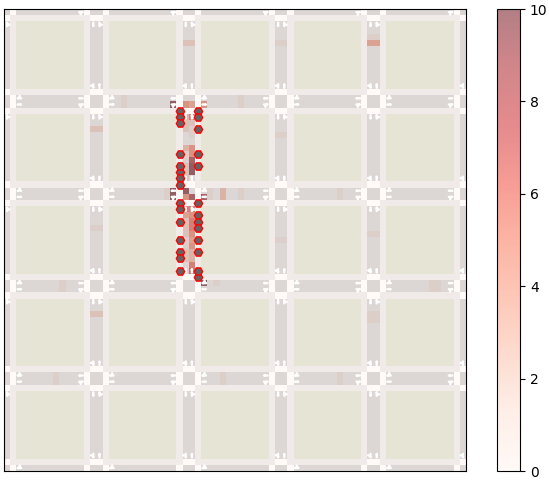}}
    \caption{Heatmaps illustrating per-cell metrics for four mobility-related phenomena. (a) Higher concentration of drivers on streets adjacent to obstructed sidewalks. (b) Reduction in vehicle speed within the obstructed or conflict-prone blocks. (c) Increased pedestrian density on sidewalks surrounding the obstacle-heavy areas. (d) Pedestrian spillover into the roadway, showing how obstacles encourage jaywalking behavior.}
    \label{fig:usecase_heatmaps}
\end{figure}



The results of the simulation are presented in Fig.~\ref{fig:usecase_heatmaps}. Panels (a) and (c) show that both drivers and pedestrians tend to redistribute toward adjacent streets and sidewalks in order to avoid the obstructed blocks, illustrating how localized infrastructure deficiencies reshape global mobility patterns. Nevertheless, agents whose destinations lie within or near the obstructed areas are forced to traverse these regions despite the increased difficulty. As seen in panel (d), pedestrians in particular are often compelled to jaywalk or move into the slower vehicle lanes when sidewalk passage is restricted, while drivers experience reduced speeds and congestion when navigating through these conflictive cells, as reflected in panel (b). This behavior highlights how obstacles not only shift traffic flow but also induce riskier interactions between agents and their environment.

\section{Conclusions}

The simulation framework developed in this study demonstrates the effectiveness of integrating agent-based modeling with a weighted A$^*$ pathfinding approach to analyze mobility dynamics in complex urban environments. By incorporating heterogeneous behavioral profiles through risk‐based weighting, the model captures realistic variability in both pedestrian and driver decisions, including risk-taking tendencies, avoidance strategies, and deviations from intended routes under obstructed conditions. This flexibility enables the simulation of nuanced behaviors that traditional macroscopic or rule-based traffic models often fail to represent.

The implementation using AgentPy allows for a high degree of configurability in population density, obstacle placement, spawning patterns, and agent capabilities. Combined with the spatial analyses performed through heatmaps, the framework supports the detection of conflict zones, including areas of reduced speed, high collision likelihood, and pedestrian spillover into vehicular lanes. These emergent patterns highlight how environmental constraints shape both individual trajectories and collective mobility outcomes.

From an applied perspective, the findings offer meaningful guidance for urban planning, pedestrian safety assessment, and infrastructure design. By simulating heterogeneous agent behaviors under varying environmental constraints, the model supports decision-making for routine mobility as well as special-event management~\cite{VILLIERS2019100052}, where temporary rerouting and crowd control are required. Additionally, the weighted path-planning approach is well suited for evacuation scenarios~\cite{pan2007agent,CABRERA2022100089}, allowing agents to adapt their routes under emergency conditions such as congestion, infrastructure damage, or increased urgency. These capabilities position the framework as a practical decision-support tool for evaluating pedestrian risk, traffic resilience, and mobility policies in both everyday and exceptional urban situations.

Overall, this work underscores the potential of multiagent simulations as a powerful tool for conflict detection, mobility assessment, and the development of smarter, safer urban spaces.

\section{Future work}

The current implementation establishes a foundational framework for modeling urban mobility through multiagent systems driven by the Weighted A$^*$ algorithm. Nonetheless, several directions can be pursued to extend the realism, adaptability, and analytical capabilities of the model.

A natural next step is the integration of real-world map data to increase spatial and contextual accuracy. Incorporating data from open sources such as OpenStreetMap or municipal geographic information systems would enable the representation of authentic street layouts, building distributions, and pedestrian pathways. This would allow simulations to be directly correlated with real environments and used to support urban planning, infrastructure assessment, or policy testing.

Another promising direction involves the use of machine learning techniques to enable adaptive and data-driven agent behavior. By combining reinforcement learning or imitation learning with rule-based decision processes, agents could learn to adjust their strategies dynamically, adapting to congestion, evolving pedestrian flows, or environmental changes. This hybrid approach could bridge the gap between deterministic pathfinding and emergent intelligent behavior.

The model can also be expanded to simulate crowd dynamics, especially relevant in high-density events or emergency evacuation scenarios. Incorporating social force models or other microscopic pedestrian interaction models would provide insight into collective movement patterns, panic behaviors, and crowd control strategies, contributing to safer and more efficient urban design.

Future iterations should also include real-time interaction between agents and urban infrastructure, such as adaptive traffic lights, crosswalk signals, and pedestrian bridges. By introducing these dynamic elements, the simulation could capture bidirectional feedback loops between agent behavior and the control policies of smart infrastructure systems, facilitating the study of cooperative and automated traffic management strategies.

In addition, a comparative analysis with alternative planning algorithms such as D*, LPA*, or reinforcement learning–based methods would provide a quantitative understanding of the trade-offs between computational efficiency, robustness, and behavioral realism. This comparison could be further enhanced by the introduction of variable weighting mechanisms, allowing agents to modify their decision priorities depending on contextual factors—for instance, increasing their weight for time efficiency when “running late” or adjusting their path when “stuck in traffic.”

Finally, extending the system to operate in real-time or online simulation modes would open opportunities for integration with sensor data or traffic monitoring systems, enabling continuous calibration of parameters and the potential deployment of digital twins for urban mobility analysis. Such extensions would transform the current model from a conceptual simulation into a versatile decision-support and experimentation platform for sustainable and intelligent urban mobility.

\section*{Appendix A: Proposed risk based modification of Weighted A$^*$}
This appendix presents a risk-aware extension of the weighted A$^*$ pathfinding algorithm used in our simulations. The modification incorporates agent-specific risk parameters into the cost function to account for behavioral variability, such as cautious or risk-seeking tendencies. By integrating these risk weights into the heuristic and movement costs, the model enables more realistic route selection and decision-making under uncertain or obstructed urban conditions.

\label{appendix:weightedastar}
\subsection*{Mathematical Formulation}

Given a graph $G = (V, E)$, the standard A$^*$ algorithm is a best-first search algorithm for the shortest path from two vertices $s, t \in V$, namely: the source and the destination. A$^*$ balances cost so far and estimated cost to goal using an evaluation function:
$$f(v)=g(v)+h(v)$$
where:
\begin{itemize}
    \item $g(v)$ is the actual cost from the start vertex to $v$. For the proposed implementation, the number of cells traveled.
    \item $h(v)$ is a heuristic estimate from any vertex $v$ to $t$. For our simulation, $h(v)$ will be calculated using the Manhattan distance: $$h(v)=|v_x-v_y|+|t_x-t_y|$$
\end{itemize}
    
Additionally, for weighted A$^*$ a weight parameter $w\geq 1$ is applied to the heuristic:
$$f(v)=g(v)+w\cdot h(v).$$
If $w=1$, the algorithm behaves as standard A$^*$, provided that $h$ is an admissible heuristic, producing optimal paths if $h$ is admissible. On the other hand, If $w>1$, the heuristic has more influence, making the search greedier, leading to shorter computation times but potentially suboptimal (riskier) paths.

\subsubsection*{Risk factor integration for urban mobility}
    For pedestrian and vehicle agents, the heuristic includes a risk penalty $r(v)$ proportional to the danger associated with traversing a cell $v$:
    $$f(v)=g(v)+w\cdot h(v)+ \alpha\cdot r(a, v)$$

where:
    \begin{itemize}
        \item $\alpha$ = risk sensitivity parameter.
        \item $r(a, v)$ = additional cost for traversing zones with risky actions (going backwards or dangerous u-turns).
    \end{itemize}

The additional cost $R(n)$ can be defined according to movement type and cell type and direction.

\subsection*{Pseudocode}
\begin{algorithm}[H]
\caption{Weighted A$^*$ Search with Manhattan Distance and Risk Penalty}
\begin{algorithmic}[1]
\Require start $s$, goal $t$, weight $w \geq 1$, risk factor $\lambda \geq 0$
\State $Q \gets$ priority queue ordered by $f(n)$
\State $g[s] \gets 0$
\State $h[s] \gets$ Dist($s$, $t$)
\State $f[s] \gets g[s] + w \cdot h[s]$
\State Push $s$ into $Q$ with priority $f[s]$
\State $from \gets$ empty map

\While{$Q$ is not empty}
    \State $current \gets$ node in $Q$ with lowest $f(n)$
    \If{$current = t$}
        \State \Return ReconstructPath($from$, $current$)
    \EndIf
    
    \ForAll{$v \in$ Neighbors(current)}
        \State $tentative\_g \gets g[current] + \text{cost}(current, v)$
        \ForAll{$a \in$ actions}
            \If{$v \notin g$ \textbf{or} $tentative\_g < g[v]$}
                \State $g[v] \gets tentative\_g$
                \State $h_{v} \gets$ Dist($v$, $t$)
                \State $f[v] \gets g[v] + w \cdot h(x) + \alpha \cdot r(a, x)$
                \State $from[v] \gets current$
                
                \If{$v$ not in $Q$}
                    \State Push $v$ into $Q$ with priority $f[v]$
                \EndIf
            \EndIf
        \EndFor
    
    \EndFor
\EndWhile
\State \Return Failure \Comment{No path found}
\end{algorithmic}
\end{algorithm}

\bibliographystyle{IEEEtran}
\bibliography{BibAgents}

\end{document}